\documentclass[citeautoscript,prx,aps,twocolumn,showpacs,showkeys,superscriptaddress,amsmath,amssymb]{revtex4}

\usepackage{graphicx}
\usepackage{psfrag}
\usepackage{epsfig}
\usepackage{color}
\usepackage{subfigure}
\definecolor{dred}{rgb}{0.7,0.0,0.0}
\usepackage{dcolumn}
\usepackage{bm}

\begin{document}

%
%
\title{Half-filled Stripes in a Hole-Doped Three-Orbital \\ Spin-Fermion Model for Cuprates}
%
%
%
\author{Mostafa Sherif Derbala Aly Hussein}
\affiliation{Department of Physics and Astronomy, University of Tennessee,
Knoxville, TN 37966, USA} 
\affiliation{Materials Science and Technology Division,
Oak Ridge National Laboratory, Oak Ridge, TN 37831, USA}


\author{Elbio Dagotto}
\affiliation{Department of Physics and Astronomy, University of Tennessee,
Knoxville, TN 37966, USA} 
\affiliation{Materials Science and Technology Division,
Oak Ridge National Laboratory, Oak Ridge, TN 37831, USA}

\author{Adriana Moreo}
\affiliation{Department of Physics and Astronomy, University of Tennessee,
Knoxville, TN 37966, USA} 
\affiliation{Materials Science and Technology Division,
Oak Ridge National Laboratory, Oak Ridge, TN 37831, USA}

\date{\today}

\begin{abstract}
{Using Monte Carlo techniques, we study a three-orbital CuO$_2$ spin-fermion model for 
copper-based high critical temperature superconductors that captures the charge-transfer properties 
of these compounds. Our studies reveal the presence of spin order
in the parent compound and, more importantly, stripe spin and charge order under hole doping. Due to the $p$-$d$ orbital hybridization,
the added holes are approximately equally distributed among the two $p$ orbitals of the oxygen atoms and the 
$d$ orbital of the copper atoms in the unit cell. In rectangular clusters of dimension $16\times 4$
{\it half-filled} stripes are observed upon hole doping, namely
when $N_h=2n$ holes are introduced in the system then $n$ stripes of length 4 are formed along the short direction. 
The original antiferromagnetic order observed in the parent compound develops
a $\pi-$shift across each stripe and the magnetic structure factor has a peak at wavevector ${\bf k}=(\pi-\delta,\pi)$ with
$\delta=2\pi N_h/N=\pi N_h/2L$, where $L=16$.
The electronic charge is also modulated and the charge structure factor is maximized at ${\bf k}=(2\delta,0)$. 
As electrons are removed from the system, intracell orbital nematicity with $\langle n_{p_x}\rangle-\langle n_{p_y}\rangle\ne 0$ develops
in the oxygen sector, as well as intercell magnetic nematicity with $\langle S^z_{{\bf i},d}(S^z_{{\bf i}+{\bf x},d}-S^z_{{\bf i}+{\bf y},d})\rangle\ne 0$ 
in the spin copper sector, in the standard notation. This occurs not only in rectangular but also in square 8$\times$8 lattices.
Overall, our results suggest that the essence of the stripe spin and charge distribution experimentally 
observed in hole-doped cuprates are captured by unbiased Monte Carlo studies of a simple hole-doped charge-transfer insulator 
CuO$_2$ spin-fermion model.}

\end{abstract}
 
\pacs{74.72.-h, 74.72.Gh, 71.10.Fd, 71.15.Dx }

\keywords{superconducting cuprates, charge-transfer insulator, multi-orbital models}
 
\maketitle

\section{Introduction} 

The parent compounds of the high critical temperature ($T_{\rm c}$) superconducting cuprates 
are known to be charge-transfer insulators (CTI)~\cite{zaanen,elbio} with a band 
structure influenced by the hybridization of the $d_{x^2-y^2}$ orbital in the copper atoms and the $p_{\sigma=x,y}$ orbitals in the oxygen atoms. 
However, due to the technical difficulty of studying  interacting many-body multiorbital Hubbard models, several of their properties, such as 
the incommensurate spin order and a tendency towards 
$d$-wave superconductivity upon doping, have been studied using simpler
single-orbital systems, such as the one-orbital Hubbard and $t-J$ models~\cite{elbio}. The use of single-orbital models 
relies on the Zhang-Rice singlet formalism 
that approximately maps a three-orbital Hubbard model into an effective $t-J$ model~\cite{ZR} and also on the photoemission 
experimental observation of a single-band Fermi 
surface~\cite{shen,allen,wells,damas}. Despite the reasonable good agreement between numerical studies on 
one- and three-orbital models~\cite{hybertsen,bacci,maria}, several authors have claimed that the multiorbital CuO$_2$ character of the 
cuprates plays a crucial role in their physics~\cite{emery,sawatzky} that cannot be neglected. 
While this issue is still being debated, it is clear that models that include the $p$-oxygen orbitals, 
in addition to the $d$-copper orbital, are more accurate and needed to study the problem of how the doped charges are distributed.
Particularly in view of the charge-transfer character of the cuprates, doped electrons primarily are located into the Cu $d$-orbitals, 
as in Mott insulators, while doped holes occupy, at least in part, the O $p$-orbitals. In fact, from
the Zaanen-Sawatzky-Allen (ZSA) paradigm~\cite{zaanen}, holes doped into a CTI should reside primarily, not only partially, in the $p$ orbitals of the oxygens. 
This is the assumption made in the Zhang-Rice approach~\cite{ZR} as well. However, recent NMR experimental results appear to indicate that the hole distribution between $p$ and $d$ 
orbitals could be material dependent~\cite{haase,haasenat}. More work is clearly needed to clarify this matter.

In addition, there is strong theoretical and experimental
interest in understanding the charge structure of the stripes observed in various hole-doped 
cuprates~\cite{tranquadastripes,birgeneau,tranq,blanco,hucker,blanco2}. Early experimental results in single-layer 
LSCO at 1/8-hole doping clearly indicated the existence of nearly static 
{\it half-filled} stripes accompanied by magnetic order commensurate with the charge stripes~\cite{tranquadastripes,birgeneau,tranq}. On the other hand, 
in bilayered YBCO, the magnetic and charge order do not appear to 
coexist~\cite{blanco,hucker,blanco2}. On the theory front, it has been recently well-established that the stripes stabilized in the ground state of the single-orbital 
Hubbard model are fully filled with holes~\cite{noack}, as opposed to half-filled. 
This appears to be a general characteristic of various single-orbital models, because it was observed in a single-orbital 
spin-fermion model for the cuprates developed by some of us in the 90s~\cite{opstripes} and, more recently, in a frustrated $t-J$ model as well~\cite{kivelson}. 
Early indications of half-filled stripes observed with density matrix renormalization group (DMRG)~\cite{dmrg} approaches in the one-orbital 
Hubbard models~\cite{stevedoug1} are 
now attributed to a finite-width effect~\cite{noack}. While there are some DMRG indications of half-filled stripes in the $t-J$ model~\cite{stevedoug} 
and more recently in the Hubbard model with additional nearest-neighbor $t'$ hopping~\cite{devereaux2}, 
the differences in the conclusions using different models and techniques underscores the need to go beyond single-orbital models to better investigate
the ground state charge and magnetic properties of hole-doped cuprates.

However, studying multiorbital models is a very challenging task. 
Magnetic stripes have been recently observed via Quantum Monte Carlo (QMC) simulations of a three-orbital CuO$_2$ 
Hubbard model~\cite{devereaux}. The simulations were performed using $8\times 8$ and $16\times 4$ clusters, as we do.
However, due to the sign problem, the studies were carried out at high temperature ($T \approx 1,000$~K), 
considerably above the regime in which the charge structure of the stripes can be studied~\cite{devereaux}. 
Also DMRG studies of the same model in $8\times 4$ clusters, smaller than discussed in our publication and using external fields at the edges
to stabilize the magnetic order,  indicate the existence of half-filled stripes~\cite{swhite}. 
Because the above mentioned QMC results are at 
high temperatures without charge order and the $8\times 4$ DMRG calculations may be affected by size effects and
need external fields for magnetic stabilization, simpler alternatives to try to capture the essence of the problem are worth investigating. 

For this reason, in this manuscript we study a recently introduced simple three-orbital spin-fermion model~\cite{3bsfm}, that captures
the properties of the charge-transfer insulating parent compound of the cuprates and that can be studied upon doping 
in a wide range of temperatures and relatively large clusters. 
This seems ideal to explore qualitatively the charge and spin properties of doped cuprates.  

Our publication is organized as follows: first, the model is described in Section~\ref{model}; then,
the magnetic and charge structures observed upon hole doping are presented in Section~\ref{results}; finally,
 Section~\ref{conclu} is devoted to the conclusions. Overall, we believe that the simple spin-fermion CuO$_2$ model is able to capture
the essence of the Cu-oxide physics with regards to the hole-doped system and its magnetic and charge properties. Extensions
of our model to larger lattices are in principle doable using the Traveling Cluster Approximation~\cite{MCMF}, as well as a study of
a wide range of temperatures, from very low to very high, the addition of quenched disorder, and the study of real-time or real-frequency 
dynamical and even d.c. transport properties. In these regards, we believe our effort opens a fertile area of research that will lead to qualitative 
progress in the study of Cu-based high-$T_{\rm c}$ superconductors and hole doped charge-transfer insulators in general.

\section{Model}\label{model} 

In the present effort, the three-orbital spin-fermion model for the cuprates~\cite{3bsfm},
which considers the 3$d_{x^2-y^2}$ Cu and 2$p_{\sigma}$ (2$p_x$ or 2$p_y$) 
orbitals of the two oxygens in the CuO$_2$ unit cell, will be studied using primarily $8\times 8$ and $16\times 4$ clusters~\cite{clusters}.
As described in our previous publication, the Hubbard repulsion at the Cu sites that splits the half-filled $d$-band is 
replaced by an effective magnetic coupling between the spin of the itinerant electrons at the $d$-orbital and 
phenomenological classical spins localized at the Cu sites. This is similar to the Mean Field Monte Carlo approximation
recently introduced \cite{MCMF}, where the local mean-field parameters in the Hartree approximation (classical variables) are coupled to
itinerant fully quantum fermions. Within this framework, 
an unbiased Monte Carlo simulation of the classical spins can be used to study
the model. In this context, there are no sign problems which means that the whole range of doping and temperatures can be explored. Since the
resulting Hamiltonian is bilinear in the fermionic operators, larger clusters than for the full multiorbital Hubbard model can be studied. 

\begin{figure}[thbp]
\begin{center}
\includegraphics[width=7cm,clip,angle=0]{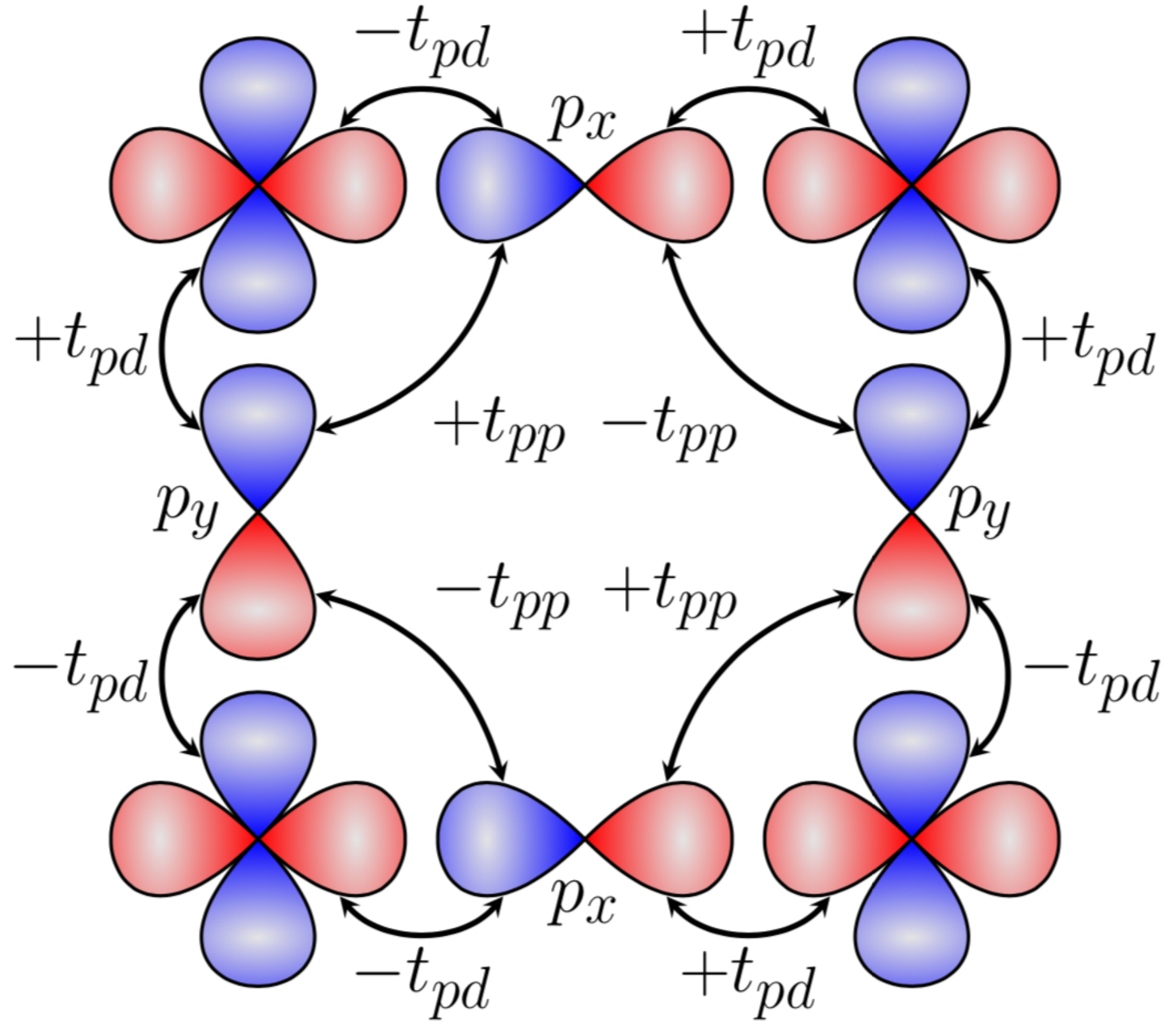}
\vskip -0.3cm
\caption{(color online) Schematic drawing of the Cu $d_{x^2-y^2}$ orbitals at the copper sites of the square lattice,
with the sign convention indicated by the colors (red for + and blue for -). The oxygen $p_{\sigma}$ orbitals 
with their corresponding sign convention are also shown, located at the Cu-O-Cu bonds.
The resulting sign convention for the $t_{pd}$ and $t_{pp}$ hoppings is also indicated.} 
\vskip -0.4cm
\label{cuo2fig}
\end{center}
\end{figure}

More specifically, the three-orbital spin-fermion (3SF) Hamiltonian 
is given by~\cite{3bsfm}
\begin{equation}
H_{\rm 3SF} = H_{\rm TB} + H_{\rm Sd} + H_{\rm AF} + H_{\rm Sp},
\label{ham}
\end{equation}
\noindent with
\begin{equation}
\begin{split}
H_{\rm TB} = -t_{pd}\sum_{{\bf i},\mu,\sigma}\alpha_{{\bf i},\mu}(p^{\dagger}_{{\bf i}+{\hat\mu\over{2}},\mu,\sigma}d_{{\bf i},\sigma}+ h.c.)-\\
t_{pp}\sum_{{\bf i},\langle\mu,\nu\rangle,\sigma}\alpha'_{{\bf i},\mu,\nu}[p^{\dagger}_{{\bf i}+{\hat\mu\over{2}},\mu,\sigma}(p_{{\bf i}+{\hat\nu\over{2}},\nu,\sigma}+p_{{\bf i}-{\hat\nu\over{2}},\nu,\sigma})+ h.c.]\\
+\epsilon_d\sum_{{\bf i}}n^d_{{\bf i}}+\epsilon_p\sum_{{\bf i},\mu}n^p_{{\bf i}+{\hat\mu\over{2}}}+\mu_e\sum_{{\bf i},\mu}(n^p_{{\bf i}+{\hat\mu\over{2}}}+n^d_{\bf i}),
\label{htb}
\end{split}
\end{equation}
\noindent where the operator $d^{\dagger}_{{\bf i},\sigma}$ creates an electron with spin $\sigma$ 
at site ${\bf i}$ of the Cu square lattice, while
$p^{\dagger}_{{\bf i}+{\hat\mu\over{2}},\mu,\sigma}$ creates an electron with spin $\sigma$ 
at orbital $p_{\mu}$, where $\mu=x$ or $y$, for the oxygen located at
${\bf i}+{\hat\mu\over{2}}$. The hopping amplitudes $t_{pd}$ and $t_{pp}$ 
correspond to the hybridizations between nearest-neighbors 
Cu-O and O-O, respectively, and $\langle\mu,\nu\rangle$ indicate O-O pairs connected by $t_{pp}$ as indicated in
Fig.~\ref{cuo2fig}. $n^p_{{\bf i}+{\hat\mu\over{2}},\sigma}$ ($n^d_{{\bf i},\sigma}$) 
is the number operator for $p$ ($d$) electrons with spin $\sigma$, while
$\epsilon_d$ and $\epsilon_p$ are the on-site energies at the Cu and O sites, respectively. 
$\Delta=\epsilon_d$ -$\epsilon_p$ is the charge-transfer gap. 
The signs of the Cu-O and O-O hoppings due to the symmetries of the orbitals is included in 
the parameters $\alpha_{{\bf i},\mu}$ and $\alpha'_{{\bf i},\mu,\nu}$ and follow
the convention shown in Fig.~\ref{cuo2fig}. The parameter values are set to
$t_{pd}=1.3$~eV and $t_{pp}=0.65$~eV. The 
on-site energy is $\epsilon_p=-3.6$~eV~\cite{hybertsen}, and thus
$\Delta=\epsilon_d-\epsilon_p$ is positive since we follow the convention $\epsilon_d=0$. 
The electron chemical potential is $\mu_e$. The remaining terms of $H_{\rm 3SF}$ are
\begin{equation}
H_{\rm Sd} = J_{\rm Sd}\sum_{{\bf i}}{\bf S_i.s_i},
\label{hSd}
\end{equation}
\noindent where ${\bf S_i}$ denotes the phenomenological localized spins at site ${\bf i}$, 
while ${\bf s_i}=d^{\dagger}_{{\bf i},\alpha}\vec\sigma_{\alpha\beta}d_{{\bf i},\beta}$ 
is the spin of the mobile $d$-electrons, with $\vec\sigma_{\alpha\beta}$ the Pauli matrices. The other
two terms are:
\begin{equation}
H_{\rm AF} = J_{\rm AF}\sum_{\langle{\bf i,j}\rangle}{\bf S_i.S_j},
\label{hAF}
\end{equation}
\noindent and 
\begin{equation}
H_{\rm Sp} = J_{\rm Sp}\sum_{{\bf i,\hat\mu}}{\bf S_i.s}_{{\bf i}+{\hat\mu\over{2}}},
\label{hSp}
\end{equation}
\noindent where $\hat\mu=\pm\hat x$ or $\pm\hat y$ and 
${\bf s}_{{\bf i}+{\hat\mu\over{2}}}=p^{\dagger}_{{\bf i}+{\hat\mu\over{2}},\mu,\alpha}\vec\sigma_{\alpha\beta}p_{{\bf i}+{\hat\mu\over{2}},\mu,\beta}$.

As mentioned above, the localized spins are assumed classical~\cite{3bsfm} which allows $H_{\rm 3SF}$ to be studied with 
the same Monte Carlo (MC) procedure widely employed before for 
the pnictides~\cite{shuhua13} and double-exchange 
manganites~\cite{manganites}. 
The values of the couplings, specifically $J_{\rm AF}=0.1$~eV, $J_{\rm Sp}=1$~eV, and $J_{\rm Sd}=3$~eV, 
were selected in our previous effort by comparing the orbital-resolved density of states (DOS)
with that of the three-orbital Hubbard model for the cuprates obtained using the variational cluster approximation on a 12-sites cluster~\cite{3bsfm}.
The calculations shown below were performed using squared $8\times 8$ and rectangular $16\times 4$ clusters~\cite{clusters} 
with periodic boundary conditions (PBC). These lattice sizes are larger than those accessible to study the
three-band Hubbard model either via quantum Monte Carlo~\cite{muramatsu,gubernatis,johnston} or via DMRG~\cite{swhite}.
During the simulation the localized spins ${\bf S_i}$ evolve using a standard Monte Carlo procedure, 
while the resulting single-particle fermionic matrix is exactly diagonalized. The simulations are performed 
at inverse temperature $\beta=(k_BT)^{-1}$ ranging from
$10$ to $800$ in units of eV$^{-1}$, equivalent to temperatures $T$ from 1200~K to 15~K~\cite{foot1}. 
In the electron representation the undoped case corresponds to 
one hole at the coppers and no holes at the oxygens, i.e.
5 electrons per CuO$_2$ unit cell (the maximum possible electronic 
number in three orbitals is 6).

\begin{figure*}
\includegraphics[width=0.8\textwidth]{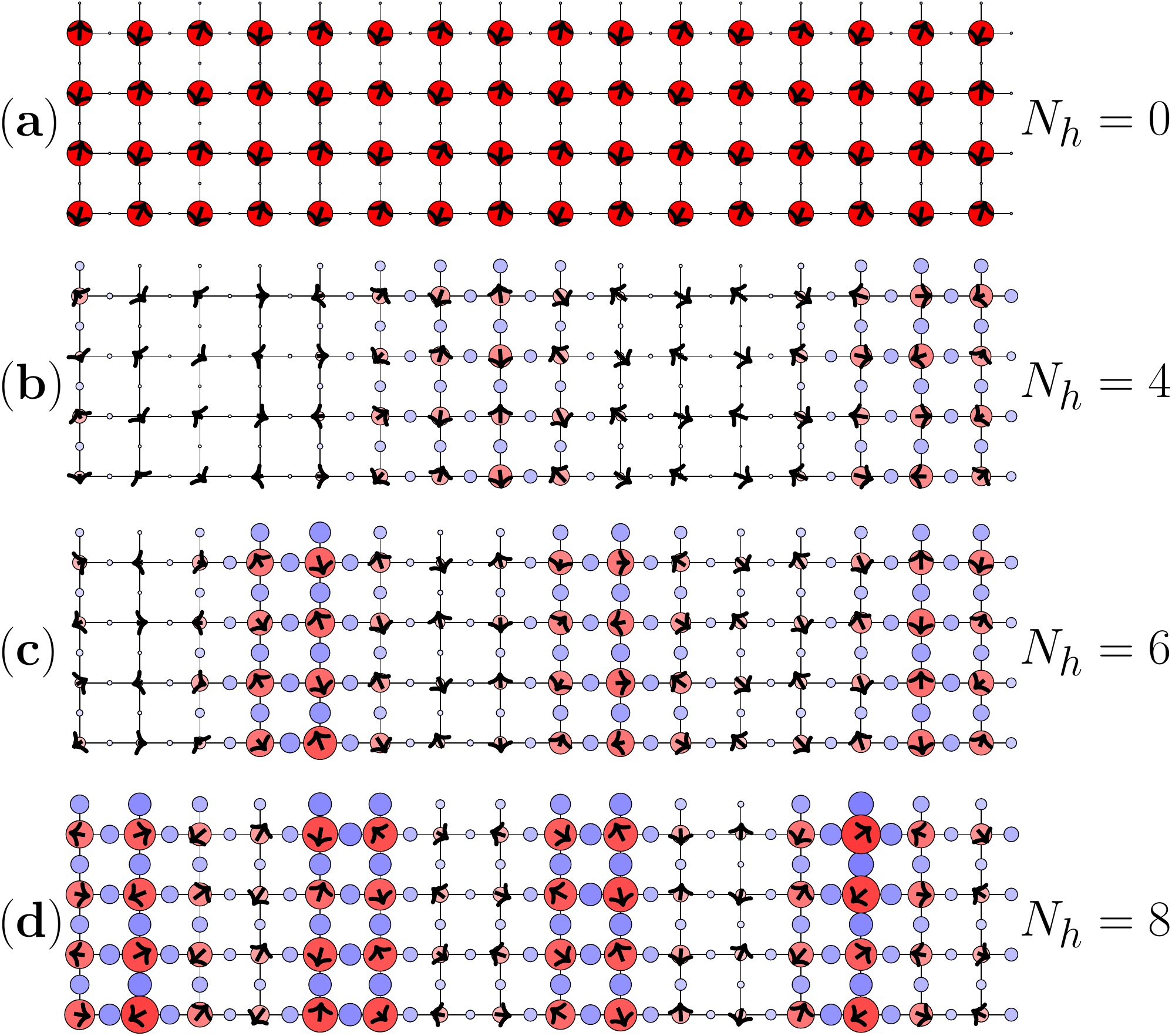}
\caption{(color online) Charge and spin configurations obtained with the spin-fermion model using $J_{\rm AF}$=0.1~eV, $J_{\rm Sp}$=1~eV, 
and  $J_{\rm Sd}$=3~eV and employing $16\times 4$ clusters at 
$\beta=800$~eV$^{-1}$ (i.e. $T \sim 15$~K) for the following electronic densities: 
(a) the undoped case with 5 electrons (i.e. 1 hole) per unit cell, with the radius of the circles proportional to the hole charge which is $n_d^h=0.82$ in 
the Cu sites and $n_p^h=0.09$ in the O sites (nearly uniform distribution) for the couplings used in our Hamiltonian; (b) results for 4 doped holes; (c) 6 doped holes; (d) 8 doped holes. 
In panels (b), (c), and (d) the radius of the circles are proportional
to the difference between the electronic density in the doped system and that in the undoped case panel (a) to better visualize the hole positions. 
The arrows in all panels are proportional to the classical spin projection in the $x-z$ plane shown.}
\label{stripes}
\end{figure*}

\section{Results}\label{results}

\subsection{Charge and spin structures}

\begin{figure}[thbp]
\begin{center}
\includegraphics[width=8.5cm,clip,angle=0]{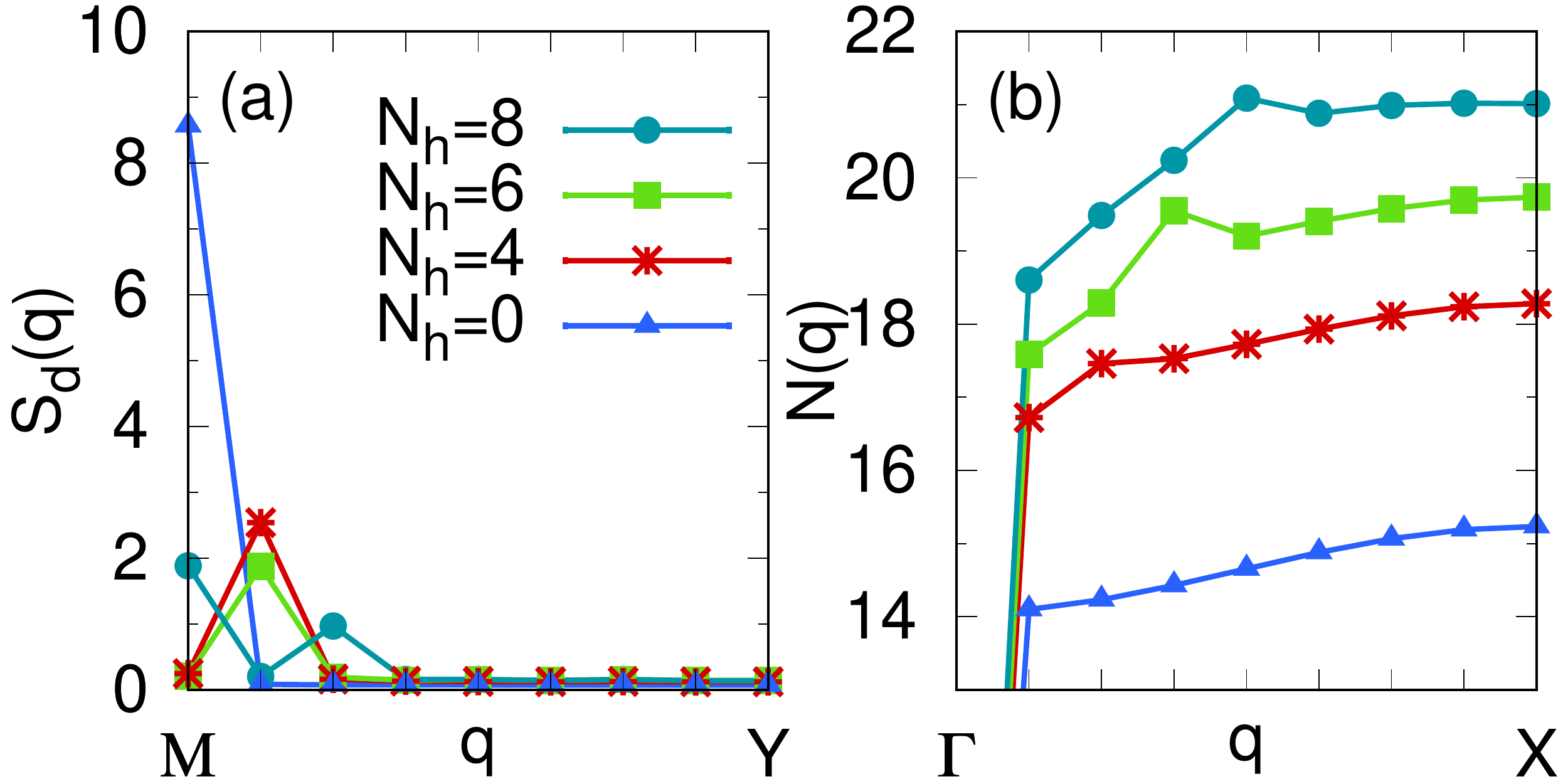}
\vskip -0.3cm
\caption{(color online) (a) The magnetic structure factor $S_d({\bf k})$ for the spin of the electrons in the Cu orbital along the M-Y direction 
in the spin-fermion model 
with $J_{\rm AF}$=0.1~eV, $J_{\rm Sp}$=1~eV, and  $J_{\rm Sd}$=3~eV, using a $16\times 4$ cluster and $\beta=800$ eV$^{-1}$ ($T \sim 15$~K) for the
number of doped holes indicated in the caption. (b) The total charge structure factor $N({\bf k})$ along the $\Gamma$-X direction 
for the same parameters as in panel (a).} 
\vskip -0.4cm
\label{sknk}
\end{center}
\end{figure}

The undoped system with 5 electrons per unit cell shows antiferromagnetically ordered localized spins and an almost 
uniform distribution of the electronic charge. For $\beta=800$~eV$^{-1}$ ($T \sim 15$~K) we found numerically that $<n_d>=1.164$ 
and $<n_{p_{\sigma}}>=1.918$, close but not identical to $<n_d>=1$ and $<n_{p_{\sigma}}>=2$ which would have been the values in the absence of 
$p-d$ hybridization. These results are virtually independent of the cluster size used~\cite{clusters}. In panel (a) of Fig.~\ref{stripes}, we display
circles which are proportional to the local hole density given by $<n_{{\bf i},\alpha}^h>=2-<n_{{\bf i},\alpha}>$ 
with $\alpha=d$ or $p_{\sigma}$ and ${\bf i}$ the site index, using a $16\times 4$ cluster at $T \sim 15$~K. 
The arrows denote the orientation of the localized spins in 
the $x-z$ plane and clearly show the staggered antiferromagnetic order that develops~\cite{footspin}. 
This magnetic order characterizes also the mobile quantum spins in the Cu, as shown by the peak 
at wavevector ${\bf k}=(\pi,\pi)$ that develops in the magnetic structure factor in panel (a) of Fig.~\ref{sknk} (triangles). 
The uniform charge distribution is indicated by the featureless charge 
structure factor $N({\bf k})$ shown in panel (b) of the same figure (triangles).

Consider now the case of doping corresponding to 4 holes. The charge is no longer uniformly distributed as shown in Fig.~\ref{stripes}~(b),
where in this panel the size of the circles is proportional to the difference between the density $n_{{\bf i},\alpha}$ and the corresponding electronic 
density in the undoped case, panel (a), to better visualize the stripes. It is clear from panel (b)
that two hole-rich stripes develop. To a good approximation, there are two holes per stripe indicating that each stripe is {\it half-filled}, as it is the case in the real
hole-doped cuprates according to neutron experiments~\cite{tranquadastripes,birgeneau,tranq,blanco,hucker,blanco2}. Figure~\ref{sknk}~(b) (crosses) 
shows that a distinct feature appears in $N({\bf k})$ at ${\bf k}=(\pi/4,0)=(2\delta,0)$ where $\delta$ indicates the displacement of 
the peak in the magnetic structure factor, that now is located at ${\bf k}=(\pi-\delta,\pi)=(7\pi/8,\pi)$ as shown in panel (a) of the figure (crosses).
The incommensuration in the quantum spins indicates the presence of $\pi$-shifts in the magnetic order across the stripes, which can be observed 
also visually in the planar 
projection of the classical spins in Fig.~\ref{stripes}~(b)~\cite{clas}.

The formation of additional half-filled stripes continues 
as more holes are added. For example, 6 (8) holes 
form 3 (4) stripes, as shown in panels (c) and (d), respectively, of Fig.~\ref{stripes}. The evolution of the magnetic and charge incommensuration $\delta$ with doping
is observed also in Fig.~\ref{sknk} where the peaks in the structure factors continue to shift. 
Notice that for 6 holes (squares) the peak in $N({\bf k})$ indicates that 
$2\delta=3\pi/8$, but since $k=3\pi/16$ is not allowed in the finite lattice used, the peak in $S({\bf k})$, 
that should be at $(13\pi/16,\pi)$, is still located at $(7\pi/8,\pi)$ (squares). For
8 holes (circles) $\delta=\pi/4$ from the peak in $N({\bf k})$ and thus, $S({\bf k})$ shows a peak at $(3\pi/4,\pi)$ which coexists with another peak at $(\pi,\pi)$. 
Notice that 8 holes corresponds to 1/8 doping in the $16\times 4$ 
cluster and the well-known $4a$ periodicity (with $a$ the lattice constant) is observed. The coexistence of the incommensurate peak 
with that at $(\pi,\pi)$ for 8 holes was found to be ubiquitous for this doping in our simulations. 
It appeared both when a random spin configuration was used as starting point of the Monte Carlo simulation 
or when an ordered spin configuration with a maximum at $(3\pi/4,\pi)$ in the spin structure factor 
was used. In both cases, ordered and disordered starting spin configuration, the simulation converged to the same final state characterized 
by 4 charge stripes and the double-peaked magnetic structure which we found already present in the snapshots. Larger clusters will be needed in order to explore 
whether the peak at $(\pi,pi)$ arises from a finite size effect.  

\subsection{Nematicity}

\begin{figure}[thbp]
\begin{center}
\includegraphics[width=8.5cm,clip,angle=0]{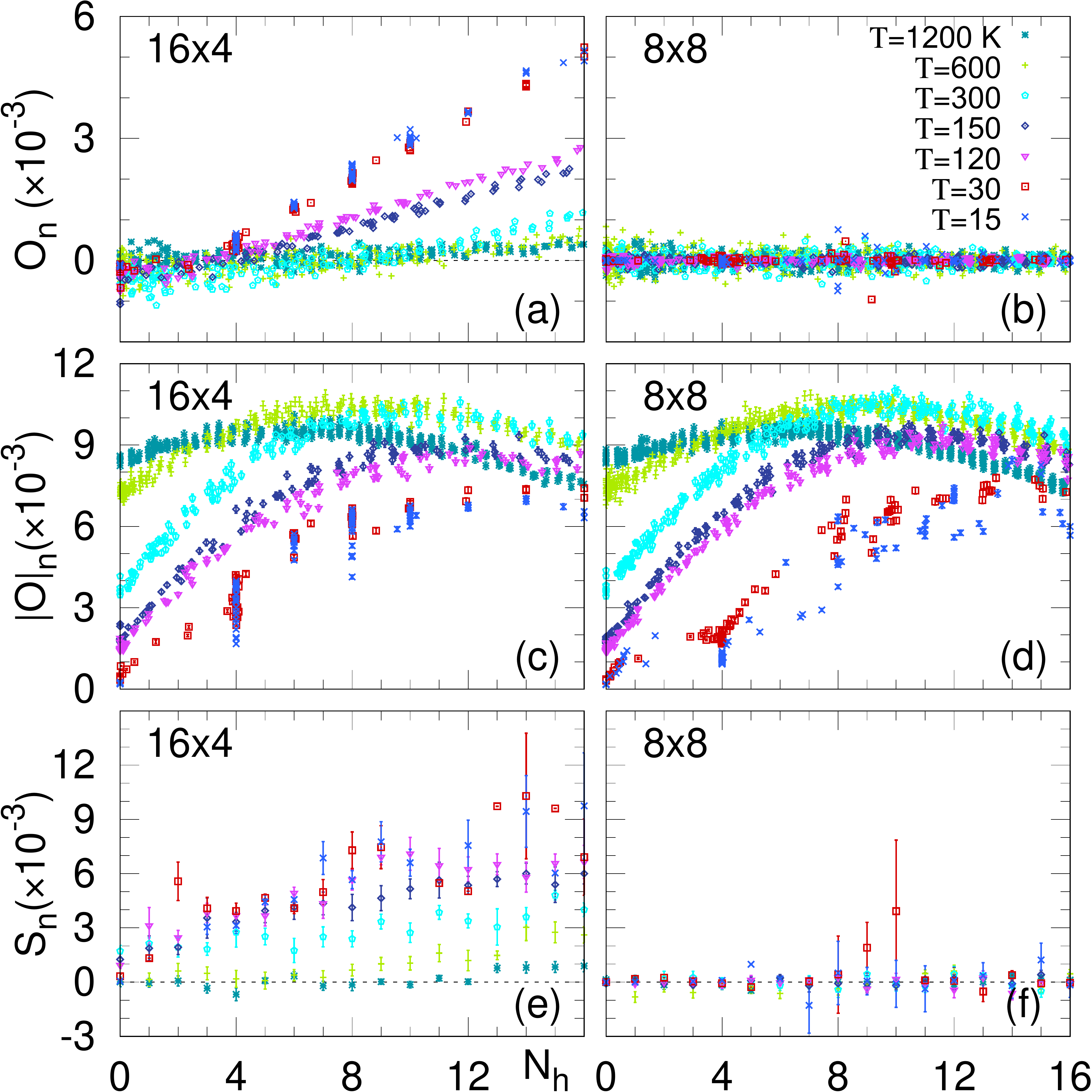}
\vskip -0.3cm
\caption{(color online) Orbital nematic order parameter $O_n=\langle n_{{\bf i},p_x} -n_{{\bf i},p_y}\rangle$ varying 
the doped number of holes at the $\beta$s 
indicated in the inset, employing a spin-fermion model with $J_{\rm AF}$=0.1~eV, $J_{\rm Sp}$=1~eV, and  $J_{\rm Sd}$=3~eV. In panel (a) 
a $16\times 4$ cluster is used. Panel (b) is the same as (a) but employing an $8\times 8$ cluster. In
(c) the nematicity is given by $|O|_n=\langle |n_{{\bf i},p_x} -n_{{\bf i},p_y}|\rangle$ using a $16\times 4$ cluster with the same parameters as panel (a).
Panel (d) is the same as (c) but in an $8\times 8$ cluster. Panel (e) same as (a) but for the spin nematic order 
parameter $S_n=\langle S^z_{{\bf i},d}(S^z_{{\bf i}+{\bf x},d}-S^z_{{\bf i}+{\bf y},d}) \rangle$. Panel (f) is the same as (e) but using an $8\times 8$ cluster.}
\vskip -0.4cm
\label{nematic}
\end{center}
\end{figure}

Together with the stripes, an interesting feature that develops with doping is $p$-orbital nematicity. In Fig.~\ref{nematic}~(a) the orbital nematic order parameter defined as
$O_n=\langle n_{{\bf i},p_x} -n_{{\bf i},p_y}\rangle$ is shown vs the number of doped holes and for values of $\beta$ ranging from 10 to 800~eV$^{-1}$
(temperatures $T$ ranging from $\sim 1,200$ to $\sim 15$~K). As expected, there is no nematicity in the undoped system. 
However, it is clear that as hole doping increases and as the temperature decreases then nematicity develops, with a larger hole occupation of the $p$ orbitals 
in the direction parallel to the stripes.
It can be argued that the nematicity is merely the result of the breaking of the rotational invariance due to 
the shape of the $16\times 4$ clusters used here. However, a non-neglibible nematic order parameter 
only develops at low temperatures and under hole doping. To further explore this issue
we evaluated the nematicity in a {\it symmetric} $8\times 8$ cluster. Here, no nematicity was observed  in $O_n$, panel (b), but this could be 
due to the coexistence of nematic regions with positive and negative values of $O_n$ switching from one another 
during the Monte Carlo time evolution, or simply a coherent
quantum mechanical superposition of both orientations.

To explore this possibility we studied the modified
order parameter $|O|_n=\langle |n_{{\bf i},p_x} -n_{{\bf i},p_y}|\rangle$. This order parameter does {\it not} change sign if the orientation
of the stripes switches from vertical to horizontal, but it becomes zero if there are no stripes.
%
In  panel (c) of Fig.~\ref{nematic} $|O|_n$ is ploted vs hole doping and at various temperatures for the $16\times 4$ cluster. 
As expected its value decreases with temperature and at $\beta =400$
and 800~eV$^{-1}$ (temperatures $T \sim 30$ and $\sim 15$~K, respectively) the data for $O_n$ in panel (a) are qualitatively reproduced (although
with different slopes). 
The Monte Carlo results for $|O|_n$ in the $8 \times 8$ cluster 
are shown in panel (d) of the figure. It is remarkable to observe that the curves are very {\it similar} to 
those for the $16\times 4$ cluster in panel (c). This clearly supports the notion that the absence of stripes on the 8$\times$8 cluster is
due to a cancellation between both orientations, with each one dominating in different regions of the system. 
Local nematicity for the square lattice with periodic boundary conditions in both directions is present 
even at the lowest temperatures reached in our numerical simulations, but combined with the results shown in 
panel (b) we can deduce that in about 50\% of the sites $n_{{\bf i},p_x} >n_{{\bf i},p_y}$ and vice versa. Namely, there is an asymmetry between the $x$ and $y$ directions. 
It is also clear that there 
is no nematicity in the undoped system at low temperature, even using rectangular clusters that in principle break the lattice rotational invariance. However,
the nematicity clearly  increases with hole doping. 

\begin{figure*}[thbp]
\begin{center}
\includegraphics[width=1.0\textwidth]{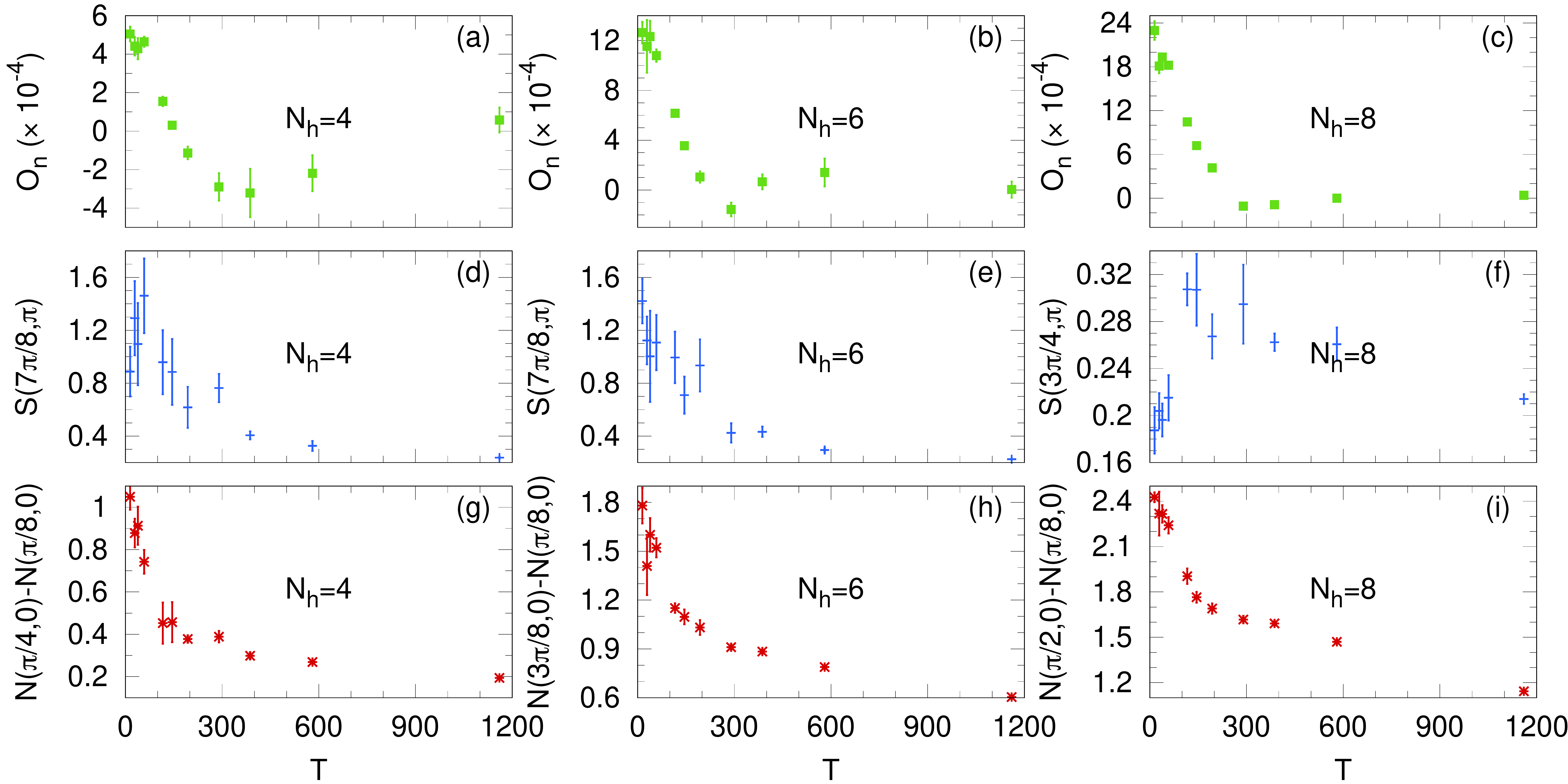}
\vskip -0.3cm
\caption{(color online) Orbital nematic order parameter $O_n=\langle n_{{\bf i},p_x} -n_{{\bf i},p_y}\rangle$ vs 
temperature $T$ for (a) 4 holes $N_h=4$, (b) 6 holes $N_h=6$, and (c) 8 holes $N_h=8$. 
We also show the magnetic structure factor $S({\bf k}_{max})$ at the wavevector where it is maximized vs 
temperature for (d) $N_h=4$ and ${\bf k}_{max}=(7\pi/8,\pi)$, (e) $N_h=6$ and ${\bf k}_{max}=(7\pi/8,\pi)$, and 
(f) $N_h=8$ and ${\bf k}_{max}=(3\pi/4,\pi)$. 
Similarly, we show an analogous analysis for the charge. Shown are the maximum value of the charge structure factor $N({\bf k}_{max})-N(\pi/8,0)$ vs 
temperature for (g) $N_h=4$ and ${\bf k}_{max}=(\pi/4,0)$, (h) $N_h=6$ and ${\bf k}_{max}=(3\pi/8,0)$, and 
(i) $N_h=8$ and ${\bf k}_{max}=(\pi/2,0)$. The results are for the spin-fermion model 
with $J_{\rm AF}$=0.1~eV, $J_{\rm Sp}$=1~eV, and  $J_{\rm Sd}$=3~eV using a $16\times 4$ cluster.}
\vskip -0.4cm
\label{multi}
\end{center}
\end{figure*}

Scanning tunneling microscopy (STM) experiments have reported intracell nematicity in the $p$ orbitals in underdoped Bi$_2$Sr$_2$CaCu$_2$O$_{8+\delta}$ 
(Bi-2212)~\cite{seamus} and in the overdoped regime of (Bi,Pb)$_2$Sr$_2$CuO$_{6+\delta}$ (Bi-2201)~\cite{hoffman}. The nematicity found 
was attributed to inequivalence in the electronic structure at the two oxygen sites within each unit cell, but the experiments could not 
disentangle whether it was of charge or magnetic origin. Our results indicate that the nematicity arises from a charge 
difference among the intracell $p_{\sigma}$ orbitals.

In addition, the previously mentioned STM experiments~\cite{seamus,hoffman} did not observe nematicity associated with the $d$ orbitals. 
However, the results of Resonant X-ray Scattering in the stripe phase of (La,M)$_2$CuO$_4$ ($M$=Sr, Ba, Eu, or Nd)~\cite{nematno} reported 
nematicity in the $d$ orbitals. Our simulations indicate that the spin 
correlations among the spin of the electrons in the $p$ orbitals are much smaller than those among the $d$ electrons 
and no magnetic nematicity in the $p$ orbitals was observed. However, 
we studied the charge correlations along the $x$ and $y$ direction for the $d$ orbital and its corresponding spin-nematic order parameter 
$S_n=\langle S^z_{{\bf i},d}(S^z_{{\bf i}+{\bf x},d}-S^z_{{\bf i}+{\bf y},d}) \rangle$.
While no nematicity was observed in the charge correlations 
we found that in the $16\times 4$ cluster the nearest-neighbor 
antiferromagnetic correlations are stronger (weaker) in the direction parallel (perpendicular) to the stripes and the anisotropy 
increases when the temperature decreases, as shown in Fig.~\ref{nematic}~(e). The corresponding results in the $8\times 8$ cluster, panel (f), do not display nematicity, but we believe that, 
as in the orbital case, panel (b), this is merely due to the equal presence of coexisting regions with
both orientations of the nematicity. After all, if the charge patterns are an equal-weight mixture of vertical and horizontal stripes, the
same has to occur for the spin textures.

\begin{figure}[thbp]
\begin{center}
\includegraphics[width=7cm,clip,angle=0]{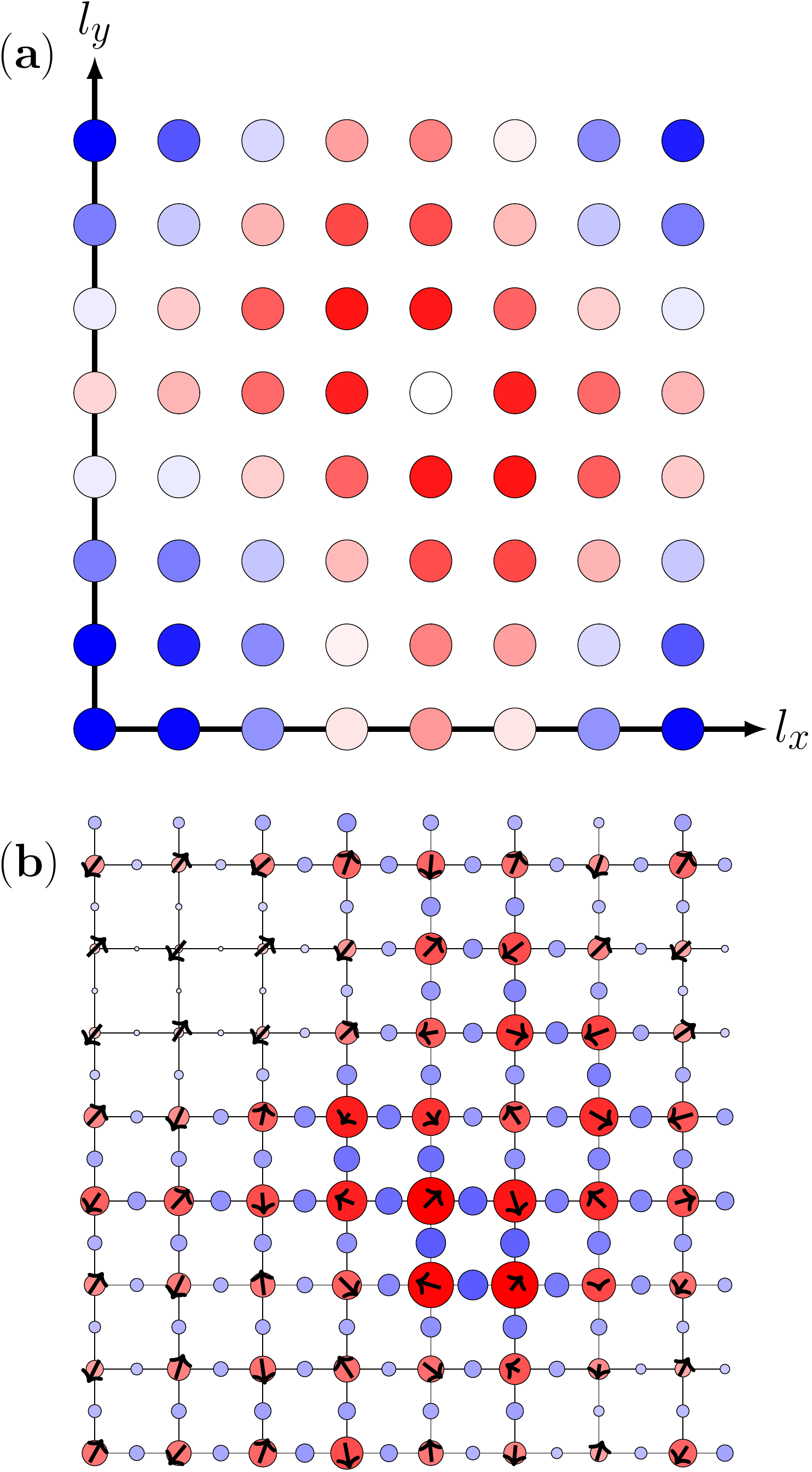}
\caption{(color online) (a) The real-space spin correlation functions 
$(-1)^{|\ell_x+\ell_y|}S^z({\bf \ell},d)$ for the electrons in the $d$ orbital using an $8\times 8$ cluster with 8 doped holes 
at $\beta=800$~eV$^{-1}$ ($T \sim 15$~K) and employing the spin-fermion model with $J_{\rm AF}$=0.1~eV, 
$J_{\rm Sp}$=1~eV, and  $J_{\rm Sd}$=3~eV. The blue tone points near (0,0) bottom left, indicate a standard spin antiferromagnetic pattern.
The red tone points indicate a spin correlation that has changed sign, namely the presence of a $\pi$-shift as it occurs in the presence of stripes. (b) Snapshot of the final configuration in a 
Monte Carlo run for the parameters in (a) showing the charge distribution and the classical spins projection in the $x-z$ plane, 
as in Fig.~\ref{stripes}. }
\vskip -0.4cm
\label{Sz}
\end{center}
\end{figure}

We have also studied how the orbital nematicity and the magnetic and charge incommensurability develop vs temperature and doping. 
In the left column of Fig.~\ref{multi} the orbital nematic order parameter $O_n$, panel (a), the magnetic structure factor 
as its maximum value ${\bf k}_{max}=(7\pi/8,\pi)$, panel (d), and the charge structure factor as its maximum value ${\bf k}_{max}=(3\pi/4,0)$ 
relative to its value at ${\bf k}=(\pi/8,0)$, panel (g), are presented for the case of 4 doped holes in the $16\times 4$ cluster. 
The three order parameters start developing at approximately the same temperature between 200~K and 300~K. For 6 doped holes the corresponding 
results appear in panels (b), (e), and (h) of the same figure and it can be observed that the three magnitudes start to increase 
at temperatures below $T\approx 300$~K. Finally, in panels (c), (f), and (i)
the results for $N_h=8$ are presented. Now the temperature below which the three order parameters start rising is lower with $T\approx 200$~K. 

These results seem to indicate that 
magnetic and charge incommensurability develop simultaneously with the nematicity. Thus, no purely isolated nematic phase is observed upon cooling. 
The presence of magnetic stripes at high temperature as reported in quantum Monte Carlo simulations of a 
three-band Hubbard model~\cite{devereaux} is not detected by our approach either.  We indeed used the approach in Ref.~\cite{devereaux}    
to understand how the cluster geometry affects the formation of stripes.  We meassured the quantum spin-spin correlations for the $d$ electrons 
$S^z({\bf \ell},d)= \langle S^z_{{\bf i},d} S^z_{{\bf i+\ell},d} \rangle$.
in real space and in panel (a) of Fig.~\ref{Sz} we display $(-1)^{|\ell_x+\ell_y|}S^z({\bf \ell},d)$ in an $8\times 8$ cluster at $\beta=800$~eV$^{-1}$ ($T \sim 15$~K) 
doped with 8 holes. 
A structure consistent with coexisting vertical and horizontal half-filled stripes
as in Ref.~\cite{devereaux} is observed: near the origin of coordinates bottom left, the blue tone points indicate a standard
staggered spin pattern, while the red tone points elsewhere indicate the presence of a $\pi$-shift in the staggered pattern as it occurs
in the presence of stripes. 
However, we have only identified these structures at low temperatures, corresponding to the 
temperatures for which the stripes are well developed
in the $16\times 4$ clusters. In addition, in panel (b) of the figure it can be seen that the charge distribution 
is also consistent with the coexistence of one horizontal and one vertical half-filled stripe.

\subsection{Total vs orbital doping}

\begin{figure}[thbp]
\begin{center}
\includegraphics[width=8.5cm,clip,angle=0]{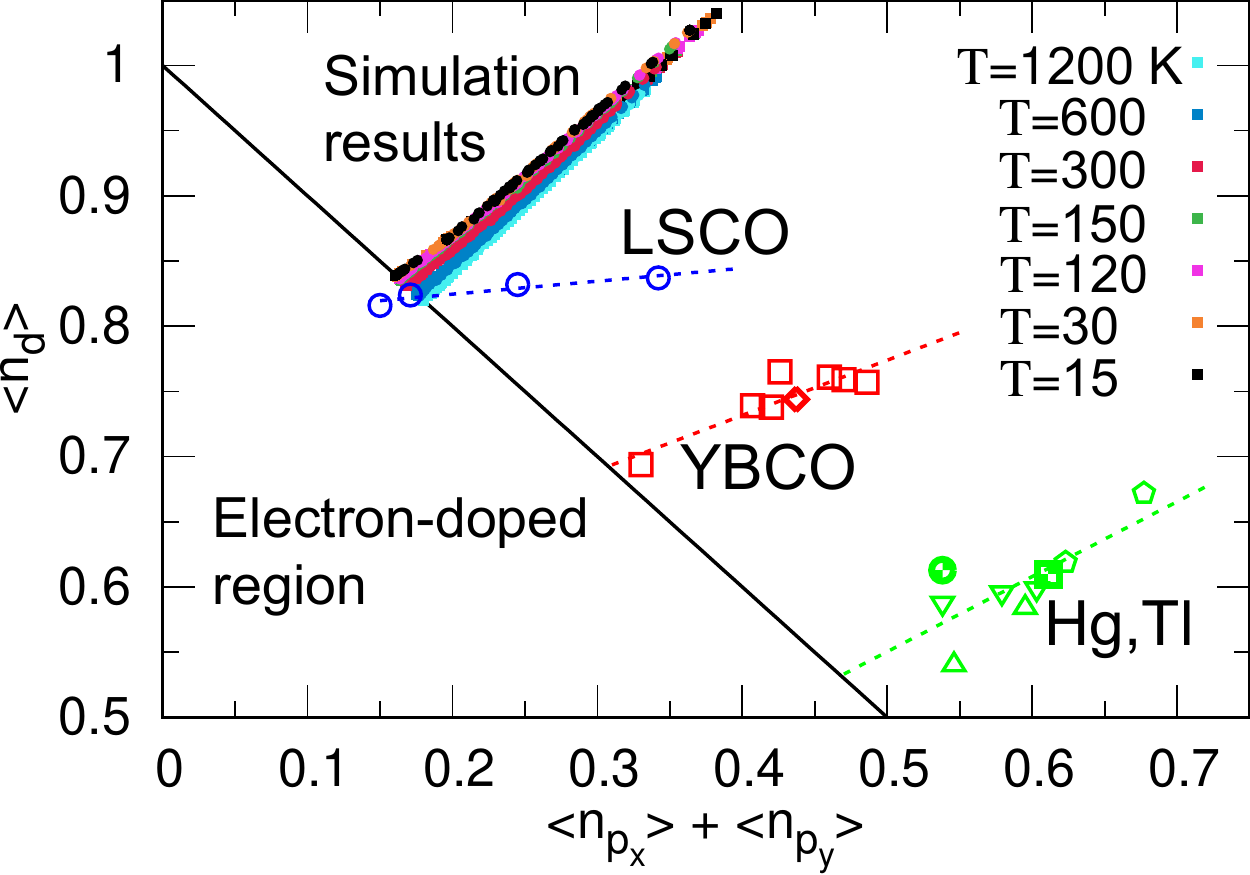}
\vskip -0.3cm
\caption{(color online) Density of holes in the $d$ orbital vs the total hole density in the $p$ orbitals for the spin-fermion model 
with $J_{\rm AF}$=0.1~eV, $J_{\rm Sp}$=1~eV, 
and $J_{\rm Sd}$=3~eV using $16\times 4$ and $8\times 8$ clusters for the temperatures indicated in the caption and for 
different hole densities, ranging from 0 holes (left) to 20 holes (right).
The solid line indicates $\langle n_d \rangle+\langle n_{p_x} \rangle+\langle n_{p_y} \rangle=1$ satisfied by the undoped system. 
Experimental results for hole doped La-214 (LSCO) (circles),
Y-124 (diamonds) and Y-123 (squares) (YBCO), and Hg- and Tl-based cuprates (green symbols) (Hg-1201, Tl-2212, Tl-2223, Tl-2201) 
were kindly provided by the authors of Ref.~\cite{haase}.
For reference, $\beta=800$~eV$^{-1}$ means $T \sim 15$~K, while $\beta=10$~eV$^{-1}$ means $T \sim 1,200$~K.} 
\vskip -0.4cm
\label{haase}
\end{center}
\end{figure}

Finally, we want to address the issue of whether the properties of the cuprates should be discussed in terms of the total doping or instead focusing on the 
local $n_d$ and $n_p$ doping as proposed in Refs.~\cite{haase,haasenat}. NMR meassurements in different superconducting cuprates indicate that the change in the 
electronic density in the $d$ and $p$ orbitals as holes are added to the system is material dependent. 
In the undoped case, with one hole per unit cell, the hole would be expected to be located at the Cu's so that the density of holes in 
the $d$ orbitals $\langle n_d^h \rangle=1$ while the density of holes in the $p$ orbitals would be $\langle n_{p_x}^h \rangle=\langle n_{p_y}^h \rangle=0$. 
However, experiments indicate that while the relationship  $\langle n_d^h \rangle+\langle n_{p_x}^h \rangle+\langle n_{p_y}^h \rangle=1$ is satisfied, 
the holes are distributed among the three orbitals in
an hybridization dependent way peculiar to each material with $\langle n_d \rangle$ ranging from 0.82 for the case of La-214 (circles in Fig.~\ref{haase}) 
to about 0.68 for Y-123 
(squares in Fig.~\ref{haase}), and finally to about 0.5 for Tl-2223 (green symbols in Fig.~\ref{haase})~\cite{haase,haasenat}. 

The results for $\langle n_d^h \rangle$ versus $\langle n_{p_x}^h \rangle+\langle n_{p_y}^h \rangle$ meassured in the spin-fermion model with 
the usual set of parameters are also plotted in
Fig.~\ref{haase} for various values of the inverse temperature $\beta$ and in $16\times 4$ and $8\times 8$ clusters. Our results indicate that the orbital 
distribution of holes has a weak temperature dependence. We observed that the
hole distribution between $d$ and $p$ electrons reproduces the experimental results for La-214 in the undoped case.  
The experimentalists also observed that the rate at which doped 
holes distribute among the $d$ and $p$ orbitals is material dependent and it is given by the slope of the curves shown in the figure. 
The slope that we observed is larger than the one obtained experimentally for the Hg and Tl compounds which, as shown in the figure, is slightly 
higher than the results for La-214~\cite{haase,haasenat}. 
We believe that a fine tuning of the parameters of the model may improve quantitatively the agreement, but it is important to notice
that qualitatively the material dependent distribution of the doped holes among the different orbitals in the unit cell is indeed 
captured by the spin-fermion model. This 
result indicates that some properties of the cuprates may be more dependent on the way in which the holes are distributed among the Cu and O orbitals 
than on the total density of doped holes.

\section{Conclusions}\label{conclu}

In this publication, we present the results of Monte Carlo studies of a phenomenological three-orbital CuO$_2$ 
spin-fermion model that captures 
the charge-transfer properties of the superconducting cuprates. The differences between a Mott and a charge transfer insulator are 
relevant upon hole doping, the regime of main focus in our present study. One of the most peculiar properties of hole-doped cuprates is the 
formation of hole half-filled stripes (one hole every two sites along the stripe)~\cite{tranquadastripes,birgeneau,tranq,blanco,hucker,blanco2,cheong,pdai,mook,tranquada}, 
as opposed to hole fully-filled stripes. 
This is a behavior that is not reproduced in the single-orbital Hubbard model~\cite{noack} and it has only been observed in three-orbital Hubbard
models using DMRG techniques in small clusters because of the numerical challenge represented by this formidable problem. Moreover,
the Quantum Monte Carlo studies of three-band Hubbard models can only be performed at temperatures above 1000~K, due to sign problems, 
where charge stripes do not exist. Thus, it is important to find simpler alternatives that capture the qualitative essence of the problem without such
computational complexity.

The present calculation
obtains for the first time {\it half-filled charge stripes} with unbiased numerical calculations of a simple spin-fermion three-orbital 
charge-transfer system. In general, it is difficult to study the stripes in square
clusters because during the Monte Carlo time evolution both vertical and horizontal stripes develop and the results represent {\it averages} in both directions. 
However, in rectangular $16\times 4$ clusters the development of half-filled stripes, accompanied by magnetic $\pi$-shifts across the stripes
is clear and properly captures the experimental results in the cuprates. In addition, we observed orbital nematicity, due to an asymmetry in the charge 
distribution between  the $p_x$ and $p_y$ orbitals
in agreement with results from STM experiments~\cite{seamus,hoffman}. Focusing on the copper $d$-orbital the 
nematicity observed with Resonant X-ray Scattering in the striped 
phase of (La,M)$_2$CuO$_4$~\cite{nematno} was also found in our analysis. 

Using 8$\times$8 clusters, and by focusing on the absolute value of the nematic
order parameter, we unveiled tendencies towards half-filled stripes even in square clusters: the average over long runs appears featureless but by using absolute values
it can be shown that there is nematicity even in square clusters. One relatively minor problem in our study
is that we found difficult to address the issue of whether the stripes are centered at the $d$ or the $p$ orbitals 
because the excess holes do not form sharp domains, as can be seen in Fig.~\ref{stripes}, but instead they have a finite width.

The correct magnetic properties are also captured by the spin-fermion model 
that displays clear tendencies towards long-range antiferromagnetic order
in the undoped case, and it also starts to develop incipient indications of incommensurability 
along $(\pi-\delta,\pi)$ and $(\pi,\pi-\delta)$ in the doped case. The coexistance of charge and magnetic order is material dependent in 
the cuprates and, in the present model, it is possible that these feature could be captured by modifications of the parameters 
in the present model. In addition, these particular features that develop upon hole doping,
likely originate in stripes, although they could also result from intertwinned orders, and they appear to require rectangular
clusters for their proper stabilization. Future work will address even larger lattices, a detailed temperature dependence,
and the influence of quenched disorder on the appearance of stripes in CuO$_2$ spin-fermion models.

\section{Acknowledgments}

All members of this collaboration were supported by the U.S. Department of Energy (DOE), 
Office of Science, Basic Energy Sciences (BES), Materials Sciences and Engineering
Division.



\begin{references}
\bibitem{clusters} Some runs in $12\times 6$ and $12\times 8$ clusters were also performed, corroborating the results obtained in the smaller clusters.
\bibitem{foot1} The temperature $T$ in Kelvin degrees is obtained using $T=11,605/\beta$.
\bibitem{footspin} The spins of the electrons in the $d$ orbital are oriented antiparallel to the classical spins.
\end{references}
\end{document}